\documentclass{article}
\usepackage{nicefrac}
\usepackage{amsfonts}
\usepackage{amssymb}
\usepackage{graphicx}
\usepackage{appendix}
\usepackage{cite}
\usepackage[intlimits]{amsmath}
\def\be{\begin{equation}}
\def\ee{\end{equation}}
\def\bea{\begin{eqnarray}}
\def\eea{\end{eqnarray}}
\begin{document}

\title{On quantum integrability of the Landau-Lifshitz model.}
\author{A. Melikyan and A. Pinzul\thanks{arsen,apinzul@iccmp.web.br.com} \\
%EndAName
\\
\emph{{International Center of Condensed Matter Physics} }\\
\emph{C.P. 04667, Brasilia, DF, Brazil} \\
\emph{and}\\
\emph{Instituto de F\'{\i}sica}\\
\emph{Universidade de S\~{a}o Paulo, 05315-970,}\\
\emph{S\~{a}o Paulo, SP, Brazil}\\
}
\date{}
\maketitle

\begin{abstract}
We investigate the quantum integrability of the Landau-Lifshitz model and
solve the long-standing problem of finding the local quantum Hamiltonian for
the arbitrary $n$-particle sector. The particular difficulty of the LL model
quantization, which arises due to the ill-defined operator product, is dealt
with by simultaneously regularizing the operator product, and constructing
the self-adjoint extensions of a very particular structure. The
diagonalizibility difficulties of the Hamiltonian of the LL model, due to
the highly singular nature of the quantum-mechanical Hamiltonian, are also
resolved in our method for the arbitrary $n$-particle sector. We explicitly
demonstrate the consistency of our construction with the quantum inverse
scattering method due to Sklyanin \cite{Sklyanin:1988s1}, and give a
prescription to systematically construct the general solution, which explains
and generalizes the puzzling results of \cite{Sklyanin:1988s1} for the
particular two-particle sector case. Moreover, we demonstrate the S-matrix
factorization and show that it is a consequence of the discontinuity
conditions on the functions involved in the construction of the self-adjoint
extensions.
\end{abstract}

\newpage
\tableofcontents
\section{Introduction}

The Landau-Lifshitz (LL) model has been the subject of great interest in
low-dimensional condensed matter physics as a model describing continuous
classical magnets (for a review see \cite{mikeska2004odm,kosevich1990ms}).
In recent years, there was a surge of interest towards the LL model in
relation with the gauge/string\ duality, where the LL\ model appeared on
both sides of the correspondence \cite{Kruczenski:2003gt, Kruczenski:2004kw,
Kazakov:2004qf, Roiban:2006yc, Stefanski:2007dp, Tirziu:2006ve,
Fradkin:1991nr}. In both one-dimensional magnetism in condensed matter
physics, and in the context of the gauge/string duality it has become clear
that the integrability plays the crucial role (for a review see \cite{Tseytlin:2004xa, Tseytlin:2003ii, Minahan:2006sk, Zarembo:2004hp}),
allowing to construct the exact solutions and revealing the rich structure
of the spectrum. Despite many years of investigation, only the classical
theory of the LL\ model has been thoroughly investigated. The classical
solitonic solutions have been found and discussed extensively in \cite{lakshmanan1977css,fogedby1980sam,fogedby1980sci,fogedby1980msp,lakshmanan1988edc,Faddeev:1987ph}. More complete classical analysis became possible after the classical
integrability was established for the isotropic case first in \cite{Takhtajan:1977rv,lakshmanan1977css}, and for the general anisotropic case
in \cite{Sklyanin:1979ll,Faddeev:1987ph}. The action-angle variables were
constructed in \cite{fogedby1980sam}, and in \cite{Zakharov:1979jc} the
classical equivalence between the LL and the non-linear Schr\"{o}dinger
(NLS) models was established, relating the flat currents of the
corresponding models by a gauge transformation. The quasi-classical spectrum
was analyzed in \cite{Jevicki:1978yv} and subsequently in \cite{fogedby1980sci}.

In contrast, the development of the quantum theory of the LL model was
affected by a number of missed subtleties and nuances, which, as a
consequence, led to the wrong quantization procedure and incorrect results 
\cite{Sklyanin:1988s1,zhao1982qch}. Let us remind the general procedure of
quantizing continuous integrable models. Apart from a few specific models,
for example, the non-linear Schr\"{o}dinger and the fermionic Thirring
models, which can be quantized directly in the continuous case by means of
the inverse scattering method as well as by coordinate Bethe ansatz, it is a
standard procedure to consider first the lattice version of the continuous
theory. This is done to deal with the ultraviolet divergences and
regularize ill-defined operator product at the same point. The quantum
Hamiltonian, and other conserved charges, can be found then by using the\
well-defined trace identities. Although there are many lattice models
corresponding to the same continuous theory \cite{Izergin:1982ry,tarasov1983lhi,tarasov1984lhi,kundu1992ciq,kundu1994slv,Korepin:1997bk}, the requirement of integrability restricts (but not eliminates completely)
the choice of the corresponding lattice model. A systematic method that in
principle is applicable for any integrable continuous model was outlined in 
\cite{Izergin:1982ry} (see also \cite{Korepin:1997bk}). In \cite{Izergin:1982ry} this program has been successfully implemented for the NLS
model and the sine-Gordon models. The difficulties with this procedure were
already emphasized in the original paper. Even for simplest NLS\ model, the
construction turned out to be quite non-trivial, and the resulting quantum
Hamiltonian had a form describing interaction of eight nearest neighbors.
For the sine-Gordon model, and in general for other continuous models, the
quantum Hamiltonian is non-local, namely, it describes the interaction which
depends on all lattice sites. An alternative method suggested in \cite{tarasov1983lhi,tarasov1984lhi} states the existence of local quantum
Hamiltonians for continuous integrable models, but its practical
construction, based on the representations of the Sklyanin algebras \cite{Sklyanin:1982tf,Sklyanin:1983ig}, is a complex and in general unresolved
problem (see also \cite{Faddeev:1992xa}). Although for the specific LL\
model, its lattice version is known, and can be obtained from the XYZ\ spin
chain, for other more complex continuous integrable models, the resulting
lattice regularized quantum Hamiltonian will in general be of a quite
complex form, if constructed following the procedure of \cite{Izergin:1982ry}. The non-locality of the Hamiltonian may be a serious barrier to deal with,
if one is interested in other subtle properties of the system. Thus, the
quantization of the continuous integrable systems is highly desirable to
carry out directly, without first passing to the lattice version.

In \cite{Sklyanin:1988s1} such program was first initiated, revealing a
number of interesting questions and nuances, associated with quantization of
the continuous integrable systems. There are several important points that
should be addressed. To begin with, the usual method of constructing the quantum
Hamiltonian and the other conserved charges, which works in lattice models
due to the well-defined expressions and a few continuous models (the NLS\
model), does not work for the LL and in general for the majority of
continuous models. The formal usage of the trace identities analogous to the
ones of the NLS\ model or the lattice models, leads to wrong results \cite{zhao1982qch}. There is, essentially, no effective or systematic method of
constructing the local conserved charges in continuous integrable models, as
the quantum corrections modify the formal ill-defined expressions, which
follow from the formal trace identities. In Sklyanin's original paper only
the action of the two-particle quantum Hamiltonian was found, which involved
some guess-work and consistency with the classical and quasi-classical
cases. The constructions also required an unusual space of quantum states in
the quantum-mechanical picture and very specific continuity properties of
the functions involved. The higher $n$-particle sector quantum Hamiltonian,
its action and the quantum states have been unknown until now. Surprisingly, this matter has not been investigated in detail, despite its obvious
importance.

There are other interesting subtleties arising in the quantization of the
LL\ model. Namely, in the anisotropic case the standard passing from the
classical to the quantum transfer matrix does not work. Instead, the $R$-matrix, as well as the monodromy matrix, are essentially guessed to satisfy
the Yang-Baxter and bilinear relations. The construction of the monodromy
matrix requires an additional spin operator, which, as a result, changes the
algebra of the spin operators, thus, giving rise to the Sklyanin algebra 
\cite{Sklyanin:1982tf,Sklyanin:1983ig}. Although the existence of such
algebras was known before for the lattice systems, in the context of the
continuous model its appearance is not very clear and remains an open
problem.

More importantly, it was realized in \cite{Sklyanin:1988s1} that there are
essentially two distinct classes of the LL\ model, corresponding to the $su(1,1)$ hyperboloid and $su(2)$ sphere cases. As it was correctly pointed
out by Sklyanin, and missed by others who attempted to quantize the LL\
model, only in the $su(1,1)$ case one may construct physically meaningful
states. In the $su(2)$ case, the scalar product turns out to be not
positively defined. Let us note, that the statement above is for the \emph{ferromagnetic} case. It is possible to construct a positively defined scalar
product for non-ferromagnentic vacuum in the $su(2)$ case \cite{albeverio1983frp}. It is well-known, however, that in the case when the
ferromagnetic vacuum does not exist, the algebraic Bethe ansatz is not
applicable, and a more sophisticated construction, similar to the
construction for the sinh-Gordon model, is needed \cite{Sklyanin:1989cf,
Sklyanin:1989cg}. This problem will be considered separately.

In this paper we consider the isotropic $su(1,1)$ LL model and construct the
desired quantum Hamiltonian for arbitrary $n$-particle sector, and its
action on the states, which we also construct in detail. We recover the
correct spectrum, and consider in detail the continuity properties of the
functions involved. For the specific case of the two-particle sector, our
results exactly reproduce Sklyanin's construction. The main point of our
method is the regularization of the Hamiltonian directly in the continuous
case. We achieve this by employing the split-point regularization, which
effectively makes the Hamiltonian non-local, and constructing the space of
quantum states, which requires a careful analysis of self-adjoint extensions
in agreement with the scalar product. We emphasize that after removing the
regularization, the Hamiltonian is local. As we show below, the quantum
mechanical Hamiltonian for the LL model yields a highly singular potential
which contains second derivatives of the delta-function. It is well known
that in the space of functions with the usual scalar product it is possible
to construct the self-adjoint extensions only up to the first derivative of
the delta-function \cite{Albeverio:1988}. More precisely, there exists a
4-parameter solution in the space of function, that contains the derivative
of the delta-function potential, and no solution exists which contains the
second derivative of the delta-function. Thus, to circumvent this barrier,
we propose to construct the self-adjoint extensions in the space of
vector-like states, which is a novel feature, not considered in the
literature previously, to the best of our knowledge. This provides an
explanation to the {\it ad hoc} solution found by Sklyanin  for the quantum mechanical Hamiltonian for the two-particle sector. Although we consider here in detail only the LL\ model as the
simplest illustrative example, the above-mentioned difficulties and
subtleties of a similar to the LL model character will exhibit themselves
also in other continuous integrable models. Thus, our method is general
enough to be applicable to a wide-range of continuous integrable models.

We also demonstrate in our method the S-matrix factorization, which is the
underlying property of quantum integrability. It is worth noting here, that
we have already considered the three-particle S-matrix factorization for the
LL\ model in our previous paper \cite{Melikyan:2008cy}. There, we
had shown that in the first non-trivial order of perturbation series, the
three-particle S-matrix is indeed factorizable into a product of the
two-particle S-matrices. Our work was based on earlier calculations of \cite{Klose:2006dd}, where the LL\ model was considered from the field theoretic
point of view, and the exact two-particle S-matrix was found by summing up
the bubble diagrams, surviving in the two-particle scattering process (see also \cite{Das:2007tb} regarding the problems with diagonalization). We
would like to emphasize the difference between the $su(1,1)$ LL model
considered here and the LL\ model considered in \cite{Klose:2006dd}. Only in
the $su(1,1)$ case one can construct a consistent quantum theory with the
ferromagnetic vacuum. It appears that the LL model considered in \cite{Klose:2006dd} actually corresponds to the $su(2)$ case. As we explained
before, the construction of positive defined metric in the space of states
corresponding to such a vacuum is mathematically challenging task, and was
considered in \cite{albeverio1983frp}. The types of excitations in the $su(1,1)$ LL model and the case of \cite{Klose:2006dd} are also different. We
do not see direct connection between the two models at this point, although
this is an interesting problem to reconstruct the results and the spectrum
of the \cite{Klose:2006dd} following the inverse scattering method and the
method we propose in this paper of regularization of the local conserved
charges together with construction of the self-adjoint extensions. It is not
surprising then, that the $n$-particle S-matrix we find here for the $su(1,1) $ LL model (which is in complete agreement with the Sklyanin's
result) is different, albeit by a coefficient, from the S-matrix found by 
\cite{Klose:2006dd}.

Our paper is organized as follows. In section \textbf{2}, we briefly review
the Landau-Lifshitz model in the context of the inverse scattering method.
In section \textbf{3.1}, we illustrate our method on the simplest two-particle case, give the regularized continuous quantum Hamiltonian,
construct the self-adjoint extensions, and derive, in complete agreement
with \cite{Sklyanin:1988s1}, the spectrum and the continuity properties of
the functions involved. In section \textbf{3.2}, we consider the general $n$-particle case and show that the regularized quantum Hamiltonian used in the
two-particle case is enough to construct the self-adjoint extensions and the
spectrum in this general case. In section \textbf{4}, we show the S-matrix
factorization as the consequence of our construction. In section \textbf{5}, we give a brief summary of our results and outline future problems.

\section{The LL model: Quantum inverse scattering method}

In this section we review, following \cite{Sklyanin:1988s1}, the main
features of the Landau-Lifshitz model in the context of the inverse
scattering method, and discuss the arising difficulties and subtleties of
this approach to quantization of the system.

The Hamiltonian for the anisotropic Landau-Lifshitz model has the following
form:
\begin{equation}
H=\frac{\epsilon}{2}\int\left[ -\left( \partial_{x}\mathbf{S},\partial _{x}\mathbf{S}\right) +4\gamma^{2}\left( \left( S^{3}\right) ^{2}-1\right) \right]  \label{Init.Ham.}
\end{equation}
where the vector $\mathbf{S}=(S^{1},S^{2},S^{3}),$ and the scalar product is
defined as $(\mathbf{S,S})\equiv\left( S^{3}\right) ^{2}-\epsilon\left(
S^{1}\right) ^{2}-\epsilon\left( S^{2}\right) ^{2}=1.$ The Poisson structure
has the form:
\begin{align}
\{S^{3}(x),S^{\pm}(y)\} & =\pm iS^{\pm}(x)\delta(x-y)  \notag \\
&  \label{P.B.} \\
\{S^{-}(x),S^{+}(y)\} & =2i\epsilon S^{3}(x)\delta(x-y)  \notag
\end{align}
where $S^{\pm}\equiv S^{1}\pm iS^{2}$. Here $\gamma$ is the anisotropy
parameter, and the choice $\epsilon=\pm1$ corresponds to the $su(1,1)$ and $su(2)$ cases correspondingly.\footnote{As we mentioned in introduction, in this article we will only consider the $su(1,1)$ case, which, unlike the $su(2)$ case, corresponds to the physically meaningful
states in the ferromagnetic vacuum, with the particular choice of the representation for the states.} As explained in \cite{Sklyanin:1988s1} the isotropic ($\gamma=0$) and anisotropic ($\gamma\neq0$) cases are essentially different,
and should be considered separately. In the former case, the conventional
inverse scattering procedure goes through without any changes - the
Yang-Baxter and bilinear equations are satisfied with the appropriate
choice of the $R$-matrix. This is in contrast to the latter case, where the
monodromy matrix and the spin operator algebra (Sklyanin algebra) have to be
modified by hand for the intertwining relation (\ref{intertwining.rel.}) to
have a solution \cite{Sklyanin:1982tf,Sklyanin:1983ig}, with the $R$-matrix
corresponding to the $XXZ$ model. The Sklyanin algebra naturally appears in
lattice systems \cite{Sklyanin:1982tf} as a consistency condition of the
intertwining relations (\ref{intertwining.rel.}). It is less clear, however,
how to derive the Sklyanin algebra directly in the continuous case, and, in
particular, in the LL model. We will consider here only the isotropic case.

The Poisson structure (\ref{P.B.}) is replaced by the commutation relations
for the $S$-operators in the standard manner:
\begin{align}
\left[ S^{3}(x),S^{\pm}(y)\right] & =\pm S^{\pm}(x)\delta(x-y)  \notag \\
&  \label{Com.rel.} \\
\left[ S^{-}(x),S^{+}(y)\right] & =2\epsilon S^{3}(x)\delta(x-y)  \notag
\end{align}
The vacuum considered here corresponds to the \emph{ferromagnetic} case:
\begin{align}
S^{3}(x)|0\rangle & =\epsilon|0\rangle  \notag \\
&  \label{fer.vac.} \\
S^{-}(x)|0\rangle & =0  \notag
\end{align}
The quantum $\mathcal{L}$-operator in the isotropic case takes the form:
\begin{equation}
\mathcal{L}(u,x)=\frac{i}{u}\left( 
\begin{array}{cc}
S^{3}(x) & -S^{+}(x) \\ 
S^{-}(x) & -S^{3}(x)
\end{array}
\right)  \label{L.operator}
\end{equation}
and the corresponding monodromy matrix is given by the expression:
\begin{equation}
T(u)=P e^{\int\limits_{-L}^{L} \mathcal{L}(u,x)dx}\equiv\left( 
\begin{array}{cc}
A(u) & B(u) \\ 
C(u) & D(u)
\end{array}
\right)  \label{monod.mat.}
\end{equation}
Here $u$ is the spectral parameter and $P$ signifies the path-ordered
exponential. In this case the bilinear relation\footnote{We use the standard notation $\overset{(1)}{T}\equiv T\otimes\boldsymbol{1}$
and $\overset{(2)}{T}\equiv\boldsymbol{1}\otimes T$} 
\begin{equation}
R(u_{1}-u_{2})\overset{(1)}{T}(u_{1})\overset{(2)}{T}(u_{2})=\overset{(2)}{T}(u_{2})\overset{(1)}{T}(u_{1})R(u_{1}-u_{2})  \label{intertwining.rel.}
\end{equation}
is satisfied with the quantum $R$-matrix given by the following form: 
\begin{equation}
R(u)=w_{a}(u)\sigma_{a}\otimes\sigma_{a}  \label{R.matrix}
\end{equation}
where the summation over the index $a=0,1,2,3$; $w_{0}(u)=u-i/2;$ $w_{1}(u)=w_{2}(u)=w_{3}(u)=-i/2.$

To construct the representations of (\ref{Com.rel.}) in the ferromagnetic
vacuum, one writes the vector in the form, analogous to other continuous
integrable models \cite{Korepin:1997bk}:
\begin{equation}
|f _{n}\rangle =\int dx_{1}...dx_{n}
f_{n}(x_{1...}x_{n})S^{+}(x_{1})...S^{+}(x_{n})|0\rangle
\label{representation}
\end{equation}
where $f _{n}(x_{1...}x_{n})$ are continuous and decreasing sufficiently
fast functions for the integral (\ref{representation}) to be well defined. A simple
calculation shows \cite{Sklyanin:1988s1} (see also the Eq. (\ref{scalprod})
below) that the scalar product $\langle g _{n}| f _{n}\rangle $ is
positively defined only for the $su(1,1)$ case, while for the $su(2)$ case,
the matrix element is indefinite. Thus, only in the $su(1,1)$ case one is
able to construct physically meaningful states in the ferromagnetic vacuum.
Therefore, one has to take $\epsilon =1$ in (\ref{Init.Ham.}-\ref{fer.vac.}).
After passing to the infinite interval, the operators in the monodromy
matrix (\ref{monod.mat.}) will satisfy the standard commutation relations,
and together with the choice of the ferromagnetic vacuum, one can apply the
well-known procedure of the algebraic Bethe ansatz to derive the spectrum
and the eigenfunctions, which have the form (see \cite{Sklyanin:1988s1} for
complete details):
\begin{equation}
|u_{1}...u_{N}\rangle =B(u_{1})...B(u_{N})|0\rangle  \label{eigenstates}
\end{equation}
In the classical and lattice models, the bilinear relation (\ref{intertwining.rel.}) guarantees the existence of the integrals of motion,
which are obtained from the generating functional $I(u)=\mathrm{Tr}\left[
T(u)\right] .$ Here, however, one faces a difficulty, which is not present
in the classical counterpart or the lattice version. Namely, in the
classical case, one can simply decompose $I(u)$ in the series :
\begin{equation}
I(u)=\underset{k}{\sum }I_{k}u^{k}  \label{the.series}
\end{equation}
to obtain the local integral of motion. For example, the classical
Hamiltonian (\ref{Init.Ham.}) can be shown to be one of the charges in the $I_{k}$ series. In the quantum case, however, the difficulty is that the
local charges, in particular the Hamiltonian (\ref{Init.Ham.}), contain
operator product at the same point, thus making the local integrals of
motion not well-defined quantities. Thus, the formally defined series (\ref{the.series}) cannot be used to obtain the integrals of motion, and, in
particular, the quantum Hamiltonian is not \emph{a priori }known. Thus,
construction of the local integrals of motion turns out to be a non-trivial
problem in the continuous quantum theory. From the field theory point of
view, this should correspond to the renormalization procedure, which to the
best of our knowledge has not yet been performed for the Landau-Lifshitz
model.

In \cite{Sklyanin:1988s1} the quantum Hamiltonian was not found, and only
the action of the (local) quantum-mechanical Hamiltonian on one- and
two-particle sectors was presented. To formulate it, it was necessary to
introduce new bosonic fields $\Psi_{n}(x),$ corresponding to the $n$-particle clusters, so that $\Psi_{n}(x)|0\rangle=0,$ satisfying the
following algebra:
\begin{equation}
\left[ \Psi_{n}(x),\Psi_{m}^{+}(y)\right] =\delta_{mn}\delta(x-y)
\label{clusters}
\end{equation}
One can then represent the $\mathbf{S}$-operators as the following $n$-particle cluster decomposition:
\begin{align}
S^{3}(x) & =s_{0}^{3}+\overset{\infty}{\underset{n=1}{\sum}} s_{n}^{3}\Psi_{n}^{+}(x)\Psi_{n}(x)  \notag \\
%&  \notag \\
S^{+}(x) & =s_{0}^{+}\Psi_{1}^{+}(x)+\overset{\infty}{\underset{n=1}{\sum}} s_{n}^{3}\Psi_{n+1}^{+}(x)\Psi_{n}(x)  \label{cluster.decomp.} \\
%&  \notag \\
S^{-}(x) & =s_{0}\Psi_{1}(x)+\overset{\infty}{\underset{n=1}{\sum}} s_{n}^{+3}\Psi_{n}^{+}(x)\Psi_{n+1}(x)  \notag
\end{align}
where $s_{0}^{3}=1;$ $s_{0}^{+}=\sqrt{2};$ $s_{n}^{3}=n;$ $s_{n}^{+}=\sqrt{(n+1)n}$ $(n\geq1).$ Using this cluster decomposition one can show that the
two-particle eigenstate (\ref{eigenstates}) has the form:
\begin{align}
|u_{1},u_{2}\rangle & =-\frac{1}{2\cos u_{1}\cos u_{2}}\left[ \int
dxe^{(p_{1}+p_{2})x}\Psi_{2}^{+}(x)\right.  \label{two.part.eigenstate} \\
& \left. +\int_{x_{x}>x_{2}}dx_{1}dx_{2}\left(
c(p_{1,}p_{2})e^{i(p_{1}x_{1}+p_{2}x_{2})}+\overline{c} (p_{1,}p_{2})e^{i(p_{1}x_{2}+p_{2}x_{1})}\right)
\Psi_{1}^{+}(x_{1})\Psi_{1}^{+}(x_{2})\right] |0\rangle  \notag
\end{align}
where
\begin{equation}
c(p_{1,}p_{2})=\frac{2(p_{1}-p_{2})+ip_{1}p_{2}}{2(p_{1}-p_{2})}
\label{c(p1p2)}
\end{equation}
Note, that in other known continuous models, solved by the coordinate Bethe
ansatz, the first term in (\ref{two.part.eigenstate}) is absent. This is the
case, for example, for the bosonic non-linear Schr\"{o}dinger and fermionic
massive Thirring models. This somewhat unusual feature of the LL\ model,
which was also discussed in \cite{Melikyan:2008cy}, will be given
explanation in the next section when constructing the self-adjoint
extensions.

As we mentioned earlier, the quantum field-theoretic Hamiltonian was not found in \cite{Sklyanin:1988s1}, but the action of the quantum-mechanical Hamiltonian on the two-particle sector was essentially guessed. To write it explicitly, it
was necessary to introduce the space spanned by the states of the following
type:
\begin{align}
|f\rangle & \equiv \left( 
\begin{array}{c}
f_{1}(x) \\ 
f_{2}(x_{1},x_{2})
\end{array}
\right)  \label{vector} \\
%&   \\
& =\left[ \int dxf_{1}(x)\Psi
_{2}^{+}(x)+\int_{x_{x}>x_{2}}dx_{1}dx_{2}f_{2}(x_{1},x_{2})\Psi
_{1}^{+}(x_{1})\Psi _{1}^{+}(x_{2})\right] |0\rangle  \notag
\end{align}
so that the function $f_{2}(x_{1},x_{2})$ is symmetric and smooth everywhere
except on $x_{1}\neq x_{2}$ line, and $f_{1}(x)=f_{2}(x,x).$ The Hamiltonian
action is defined as follows:
\begin{equation}
H_{2}|f\rangle =\left( 
\begin{array}{c}
2\left( \partial _{x_{1}}-\partial _{x_{2}}\right)
f_{2}|_{x_{2}=x_{1}-\epsilon }^{x_{2}=x_{1}+\epsilon }-\partial
_{x}^{2}f_{2}(x,x) \\ 
\\ 
-\left( \partial _{x_{1}}^{2}+\partial _{x_{2}}^{2}\right) f_{2}(x_{1},x_{2})
\end{array}
\right)  \label{Hamiltonian.action}
\end{equation}
It is not difficult to check that the above action of the Hamiltonian $H_{2}$
is Hermitian with respect to the following scalar product:
\begin{equation}
\left\Vert f\right\Vert ^{2}=\frac{1}{4}\int dx\left\vert f_{1}\right\vert
^{2}+\int_{x_{x}>x_{2}}dx_{1}dx_{2}\left\vert f_{2}(x_{1},x_{2})\right\vert
^{2}  \label{scalar.product}
\end{equation}
One can check, that the solution to (\ref{vector}-\ref{scalar.product})
leads exactly to the two-particle state (\ref{two.part.eigenstate}).

Although this form of the action was guessed by Sklyanin for the
two-particle sector, its origin and the general $n$-particle Hamiltonian
action and the corresponding space of states were unknown. In the next
section both problems will be resolved, the local quantum Hamiltonian will
be proposed, and the quantum states will be constructed.

\section{Quantum Hamiltonian and self-adjoint extensions}

As we discussed in the introduction, the construction of the local quantum
Hamiltonian is a complicated task in general. The standard procedure of
putting the continuous theory on the lattice to regularize the ultraviolet
divergences leads generally to non-local Hamiltonians, and only the
existence of the local form can be proven, while construction in practice is
a complicated and unresolved problem. On the other hand, only a few continuous
integrable models (NLS, massive Thirring model) allow direct quantization of
the system by coordinate Bethe ansatz, without using the inverse scattering
method. The essential difference of the LL model from the NLS or Thirring
models is the presence of more severe singularities in the quantum
mechanical Hamiltonian. Indeed, in the NLS\ model, the quantum mechanical
interaction is described by the $\delta (x)$ potential, while in the case of
the LL\ model the interaction is highly singular and is proportional to $\partial _{x}\partial _{y}\delta (x-y).$ The standard procedure to deal
correctly with such singular potentials is to construct self-adjoint
extensions. Even though for the NLS\ and Thirring models the problem was
solved without constructing self-adjoint extensions, in general this is not
correct.

The construction of self-adjoint extensions for the LL model is, however,
immediately bounded by the following fact. It is known that in the space of
functions with the usual scalar product it is possible to construct the
self-adjoint extensions only up to the first derivative of the
delta-function \cite{Albeverio:1988}, namely, there exists a four-parameter extension in the space of function, that contains the derivative of the
delta-function potential. Since for the LL\ model the interaction is of the
second order derivative of the delta-function, the above statement means
that one has to construct a different scalar product in the new space of
function. In the next section we will present this construction in detail
first for the two-particle case, before generalizing our analysis for the $n$-particle case. We will derive the Sklyanin's result (\ref{vector}-\ref{scalar.product}) and find its $n$-particle extensions. To do this, we will
also propose the regularized continuous quantum Hamiltonian that will be checked to give correct results for any $n$-particle sector.

\subsection{Two-particle sector}

It is easier to demonstrate the idea of our method on the two-particle case,
and in the more general case, considered in the next section, the
complications are only of the technical character. Let us begin by writing
down the regularized quantum Hamiltonian, corresponding to (\ref{Init.Ham.}). It is not difficult to see, that the direct application of the
Hamiltonian~(\ref{Init.Ham.}) to the two-particle state (\ref{representation}) 
\begin{equation}
|f_{2}\rangle =\int dxdyf_{2}(x,y)S^{+}(x)S^{+}(y)|0\rangle
\label{2-part.rep}
\end{equation}
leads to undefined singular expressions of the type $\partial _{x}^{2}\delta
(x)|_{x=0}.$ The idea is to regularize the continuous Hamiltonian by the split-point method. Namely, we take our quantum Hamiltonian to be of the
form:
\begin{equation}
H_{Q}=\underset{\varepsilon \rightarrow 0}{\lim }H_{\varepsilon }
\label{Quant.Ham.}
\end{equation}
where
\begin{equation}
H_{\varepsilon }=\frac{1}{2}\int dudvF_{\varepsilon }(u,v)\left[ -\partial
_{u}S^{3}\partial _{v}S^{3}+\partial _{u}S^{+}\partial _{v}S^{-}+\partial
_{u}\partial _{v}\left( S^{3}(u)\delta (u-v)\right) -\partial _{u}\partial
_{v}\delta (u-v)\right]  \label{Quant.Ham2.}
\end{equation}
Here the function $F(u,v)$ is any smooth function, depending on some
parameter $\varepsilon ,$ so that 
\begin{equation}
\underset{\varepsilon \rightarrow 0}{\lim }F_{\varepsilon }(u,v)=\delta (u-v)
\label{F(u,v)}
\end{equation}
We assume that $F(u,v)$ decreases rapidly enough to make (\ref{Quant.Ham2.})
well-defined. Let us emphasize, that although the intermediate Hamiltonian (\ref{Quant.Ham2.}) is essentially non-local, we will remove the
regularization (\ref{F(u,v)}) only after computations, thus, making the
theory local. The first two terms in the limit (\ref{F(u,v)}) will go to the
classical Hamiltonian (\ref{Init.Ham.}), and the last term of (\ref{Quant.Ham2.}) is introduced to remove the infinities.\footnote{Note, that this term is enough to remove infinities in all $n$-particle
sectors simultaneously (see the Eq. (\ref{inter1})).} It is not difficult to
show the following properties of the Hamiltonian $H_{Q}$: 
\begin{align}
H_{Q}|0\rangle & =0 \label{Hq.prop} \\
%&   \\
H_{Q}|f_{1}\rangle & =\int dx\left[ -\partial _{x}^{2}\right] f_{1}(x)S^{+}(x)|0\rangle  \notag
\end{align}
where $|f _{1}\rangle $ is the one-particle state 
\begin{equation}
|f_{1}\rangle =\int dx f_{1}(x)S^{+}(x)|0\rangle  \label{1.part}
\end{equation}
Thus, for the one-particle state we recover the correct solution $f_{1}(x)\sim \exp (ipx),$ with the energy $E_{1}=p^{2}.$

Let us consider now the more complex two-particle case. One can show that 
\begin{align}
H_{Q}|f_{2}\rangle & =-\int dxdy\left( \partial _{x}^{2}+\partial
_{y}^{2}\right) f_{2}(x,y)S^{+}(x)S^{+}(y)|0\rangle  \label{Ham.on.2part.} \\
%&   \\
& +\int dx\left[ \left( \partial _{x}-\partial _{y}\right)
f_{2}(x,y)|_{x=y+\epsilon }^{x=y-\epsilon }-\partial _{x}\partial
_{y}f_{2}(x,y)|_{x=y}\right] S^{+}(x)S^{+}(x)|0\rangle  \notag
\end{align}
Thus, for (\ref{2-part.rep}) to be an eigenstate of $H_{Q},$ we must require
the following matching condition:
\begin{equation}
\left( \partial _{x}-\partial _{y}\right) f_{2}(x,y)|_{x=y+\epsilon
}^{x=y-\epsilon }-\partial _{x}\partial _{y}f_{2}(x,y)|_{x=y}=0
\label{matc.cond1}
\end{equation}
We note here, that in the more general case for the $n$-particle sector,
these matching conditions, resulting after the action of the quantum
Hamiltonian (\ref{Quant.Ham.}) on the $n$-particle state, will remain the
same, with the obvious change $x\rightarrow x_{i}$ and $y\rightarrow x_{j}$.
After some straightforward algebra, we can derive the action of the quantum
Hamiltonian (\ref{Quant.Ham.}) on the $n$-particle state:
\begin{eqnarray}
H_{Q}|f_{n}\rangle &=&-\int d\overrightarrow{x}\left( \Delta f(\overrightarrow{x})\right) \overset{n}{\prod_{i=1}}S^{+}(x_{i})|0\rangle
\label{inter1} \\
&&+\sum_{i>j}\int \prod_{k\neq j}dx_{k}\left( \left( \partial _{j}f(\overrightarrow{x})-\partial _{i}f(\overrightarrow{x})\right)
|_{x_{j}=x_{i}-\epsilon }^{x_{j}=x_{i}+\epsilon }+\partial _{i}\partial
_{j}f(\overrightarrow{x})|_{x_{i}=x_{j}}\right) \overset{n}{\prod_{i=1}}S^{+}(x_{i})|_{x_{i}=x_{j}}|0\rangle  \notag
\end{eqnarray}
From here we immediately derive the general matching conditions:
\begin{equation}
-\left( \partial _{j}f(\vec{x})-\partial _{i}f(\vec{x})\right)
_{x_{j}=x_{i}-\epsilon }^{x_{j}=x_{i}+\epsilon }=\partial _{j}\partial _{i}f(\vec{x})|_{x_{i}=x_{j}}\ ,\ \ \forall i>j  \label{matching2}
\end{equation}
This is, in fact, the reason for the S-matrix factorization, as we will see
below. We also emphasize, that the quantum Hamiltonian (\ref{Quant.Ham.})
does not acquire any further corrections in the higher $n$-particle sectors.

We will show now, that the matching condition (\ref{matc.cond1}) together
with the equation following from (\ref{Ham.on.2part.})
\begin{equation}
-\left( \partial _{x}^{2}+\partial _{y}^{2}\right) f_{2}(x,y)=\left(E_{2}\right) f_{2}(x,y)
\label{solution1}
\end{equation}
where $E_{2}$ is the energy of the two-particle state, recovers the Sklyanin's solution (\ref{two.part.eigenstate}). We will first
construct the space on which the quantum-mechanical Hamiltonian acts.

Let us consider a space $V$ generated by the vectors of the form 
\begin{equation}
\Psi=\left( 
\begin{array}{c}
f_{1}(x) \\ 
f_{2}(x,y) \\ 
\end{array}
\right)  \label{vector1}
\end{equation}
where the function $f_{1}(x)$ is determined by $f_{2}(x,y)$ and, possibly,
its derivatives at $x\rightarrow y$. The actual form of $f_{1}(x)$ will be
fixed later. For a given non-negative number, $\alpha$, we define a scalar
product on $V$ as follows: 
\begin{equation}
\langle\Phi|\Psi\rangle=\alpha\int_{-\infty}^{\infty}g_{1}^{\ast}(x)f_{1}(x) \ dx+\iint_{x\neq y}g_{2}^{\ast}(x,y)f_{2}(x,y)\ dxdy  \label{scalar_prod}
\end{equation}
where 
\begin{equation}
\Psi=\left( 
\begin{array}{c}
f_{1}(x) \\ 
f_{2}(x,y) \\ 
\end{array}
\right) \mbox{\ and\ }\Phi=\left( 
\begin{array}{c}
g_{1}(x) \\ 
g_{2}(x,y) \\ 
\end{array}
\right)  \label{2vectors}
\end{equation}
We require the function $f_{1}(x)$ to belong to $\mathcal{L}^{2}(\mathbb{R},dx)$ and $f_{2}(x)$ to $\mathcal{L}^{2}(\mathbb{R}^{2}/\{x=y\},dxdy)$.
Further conditions on $f_{1}(x)$ and $f_{2}(x,y)$ will be imposed later.

Now we define the operator, $\hat{H}$, on $V$ with the following properties:

i) acting on $f_{2}(x)$ it is simply the Laplacian $-\triangle \equiv
-\partial _{x}^{2}-\partial _{y}^{2}$ everywhere in $\mathbb{R}^{2}/\{x=y\}$, i.e. 
\begin{equation}
\hat{H}\Psi =\left( 
\begin{array}{c}
\hat{h}f_{1}(x) \\ 
-\triangle f_{2}(x,y) \\ 
\end{array}
\right)  \label{Ham}
\end{equation}
with some operator $\hat{h}$ to be determined later;

ii) it is Hermitian with respect to the scalar product (\ref{scalar_prod}),
i.e. 
\begin{equation}
\langle\Phi|\hat{H}|\Psi\rangle= \langle\hat{H}\Phi|\Psi\rangle  \label{herm}
\end{equation}

Using 
\begin{equation}
g^{\ast}(x,y)\triangle f(x,y)=\partial_{i}(g^{\ast}(x,y)\partial
_{i}f(x,y))-\partial_{i}((\partial_{i}g(x,y))^{\ast}f(x,y))+(\triangle
g(x,y))^{\ast}f(x,y)  \notag
\end{equation}
where $i=\{x,y\},$ and not assuming continuity neither functions nor their
derivatives at $x=y,$ we have 
\begin{align}
& \iint_{x\neq y}g_{2}^{\ast}(x,y)\triangle f_{2}(x,y)\ dxdy=\iint_{x\neq
y}(\triangle g_{2}(x,y))^{\ast}f_{2}(x,y)\ dxdy  \notag \\
& -\int_{-\infty}^{\infty}dy\ \left[ g_{2}^{\ast}(x,y) \partial_{x}f_{2}(x,y)-(\partial_{x}g_{2}(x,y))^{\ast}f_{2}(x,y)\right]
_{x=y-\epsilon }^{x=y+\epsilon}  \notag \\
& -\int_{-\infty}^{\infty}dx\ \left[ g_{2}^{\ast}(x,y) \partial_{y}f_{2}(x,y)-(\partial_{y}g_{2}(x,y))^{\ast}f_{2}(x,y)\right]
_{y=x-\epsilon }^{y=x+\epsilon}  \label{partint}
\end{align}

Now, to be more specific, we are going to impose some conditions on $f_{2}(x,y)$. We require it to be continuous at $x=y$.\footnote{The other possibilities are also interesting but we consider the one that is
relevant for our problem.} In this case Eq.(\ref{partint}) simplifies to
become 
\begin{align}
& \iint_{x\neq y}g_{2}^{\ast}(x,y)\triangle f_{2}(x,y)\ dxdy=\iint_{x\neq
y}(\triangle g_{2}(x,y))^{\ast}f_{2}(x,y)\ dxdy  \notag \\
& +\int_{-\infty}^{\infty}dx\ \left\{ g_{2}^{\ast}(x,x)\left[ \partial
_{x}f_{2}(x,y)-\partial_{y}f_{2}(x,y)\right] _{y=x-\epsilon}^{y=x+\epsilon
}\right.  \notag \\
& \left. -\left[ (\partial_{x}g_{2}(x,y))^{\ast}-(\partial_{y}g_{2}(x,y))^{\ast}\right] _{y=x-\epsilon}^{y=x+\epsilon}f_{2}(x,x)\right\}
\label{partint1}
\end{align}
It is obvious that the first line in (\ref{partint1}) has needed for
Hermiticity form while the second and third lines should be compensated by $\hat{h}$ (see Eq.(\ref{Ham})) in such a way that the full $\hat{H}$ would
become Hermitian. Before proceeding, we must fix the relation between $f_{1}(x)$ and $f_{2}(x,y)$. Looking at the one-dimensional integral in (\ref{herm}), one sees that a natural choice is $f_{1}(x)=f_{2}(x,x)$ (which is
possible after we required continuity of $f_{2}(x,y)$). Now it is not
difficult to see that the following form of $\hat{h}$ 
\begin{equation}
\hat{h}f_{1}(x):=\frac{1}{\alpha}\left[ \partial_{x}f_{2}(x,y)-\partial
_{y}f_{2}(x,y)\right] _{y=x-\epsilon}^{y=x+\epsilon}+\hat{h}_{1}f_{1}(x)\ ,
\label{h.hat}
\end{equation}
where $\hat{h}_{1}$ is \textit{any} Hermitian in $\mathcal{L}^{2}(\mathbb{R},dx)$, does the job.

Thus, we see that the requirement for $\hat{H}$ to be Hermitian (or rather
symmetric) with respect to the scalar product (\ref{scalar_prod}) does not
fix the Hamiltonian completely even after we had chosen some conditions on
the components of $|\Psi\rangle$. But, in fact, we still have to use one
consistency condition: the image of $|\Psi\rangle$ under the action of $\hat{H}$ should belong to the same class of vectors, namely the first component
should be related to the second one 
\begin{equation}
\hat{h}f_{1}(x):=\frac{1}{\alpha}\left[ \partial_{x}f_{2}(x,y)-\partial
_{y}f_{2}(x,y)\right] _{y=x-\epsilon}^{y=x+\epsilon}+\hat{h}_{1}f_{1}(x)\equiv-\triangle f_{2}(x,y)|_{x=y}  \label{gluing}
\end{equation}
This put some constraints on the form of $\hat{h}_{1}$ as well as imposes
some conditions on the behavior of $f_{2}(x,y)$ at $x=y$.

For the LL model, it is not difficult to show from (\ref{2-part.rep}), that
the coefficient in the scalar product (\ref{scalar_prod}) $\alpha =1/2$, and
using the Eq. (\ref{matc.cond1}) we find that $\hat{h}_{1}=-\partial
_{x}^{2}.$ Thus, we obtain from (\ref{Ham}):
\begin{equation}
H_{2}|f\rangle =\left( 
\begin{array}{c}
2\left( \partial _{x_{1}}-\partial _{x_{2}}\right)
f_{2}|_{x_{2}=x_{1}-\epsilon}^{x_{2}=x_{1}+\epsilon}-\partial _{x}^{2}f_{2}(x,x) \\ 
\\ 
-\left( \partial _{x_{1}}^{2}+\partial _{x_{2}}^{2}\right) f_{2}(x_{1},x_{2})
\end{array}
\right)  \label{main.formula}
\end{equation}
This is exactly the formula (\ref{Hamiltonian.action}) guessed by Sklyanin.

\subsection{$n$-particle sector}

Let us consider a space of vectors of the form 
\begin{equation}
|f_{n}\rangle =\int d^{n}\vec{x}\ f(\vec{x})\ S^{+}(x_{1})\cdots
S^{+}(x_{n})|0\rangle  \label{wavefunct}
\end{equation}
Then it is not difficult to calculate a `scalar product' in this space. Let $\{X_{m}\}$ be a partition of the set $\{x_{i}\}$: 
\begin{equation*}
\bigcup_{m=1}^{M_{P}}X_{m}=\{x_{i}\}\mbox{ and }X_{m}\bigcap X_{n}=\delta
_{mn}X_{m}
\end{equation*}
and let $t_{m}$ be a `collective' coordinate for all $x_{i}\in X_{m}$. Then
the resulting `scalar product' is 
\begin{equation}
\langle g_{n}|f_{n}\rangle =\sum_{partitions}\epsilon ^{n-M_{p}}C_{P}\int
d^{M_{P}}t\ g^{\ast }(\vec{x})f(\vec{x})|_{\{x_{i}\in X_{m}\}=t_{m}}
\label{scalprod}
\end{equation}
Here $C_{P}$ are some combinatorial factors, which are positive. From here
it is obvious that this `scalar product' will be an actual scalar product
only for the non-compact case of $su(1,1)$, where $\epsilon =1$. From now on
we concentrate on this case postponing consideration of the $su(2)$ model to
future work. It is instructive to write down $n=2$ and $n=3$ cases
explicitly: 
\begin{align}
& n=2:\ \ \langle g_{2}|f_{2}\rangle =4\left[ 2\int dxdy\ g^{\ast
}(x,y)f(x,y)+\int dx\ g^{\ast }(x,x)f(x,x)\right]  \label{2part} \\
& n=3:\ \ \langle g_{3}|f_{3}\rangle =8\left[ 6\int dxdydz\ g^{\ast
}(x,y,z)f(x,y,z)\right.  \notag \\
& \ \ \ \ \ \ \ \ \ \ \ \ \ \ \ \ \ \ \ \ \ \left. +(3\cdot 3)\int dxdy\ g^{\ast }(x,x,y)f(x,x,y)+\int dx\ g^{\ast
}(x,x,x)f(x,x,x)\right]  \label{3part}
\end{align}
While (\ref{2part}) is exactly the scalar product used in \cite{Sklyanin:1988s1} (see also the Eq. (\ref{scalar_prod})), the formula (\ref{3part}) is presented here to explicitly demonstrate the first non-trivial case.

There are a couple of useful interpretations of (\ref{scalprod}). The first
one is along the lines of \cite{Sklyanin:1988s1}. Namely, one can think of (\ref{wavefunct}) as a vector-function with (\ref{scalprod}) as a natural
scalar product. Even though this is a useful interpretation, mostly because this
nicely fits into the cluster picture of \cite{Sklyanin:1988s1}, the more
mathematically rigorous one is as follows. Namely, it is a scalar product in $L_{\mu
}(R) $, where the completeness is defined with the help of the
Riemann-Stieltjes integral, rather than just the Riemann one. Here, $\mu $ is the
'measure' for this integral, which formally could be written as 
\begin{equation}
\int f(\vec{x})d^{n}\mu (\vec{x})\ ,  \label{unn1}
\end{equation}
where $\mu $ now is not required to be a smooth function. The only
requirement is that $f$ and $\mu $ do not have discontinuity at the same
points, which is the case - our function is continuous. And that is why it
is that hard to define the self-adjoint operator - one needs to be extremely
careful at the points, where the measure is discontinuous. The measure $\mu $ is very
easy to read from the scalar product. In particular, for the case $n=2$, Eq. (\ref{2part}), we have (formally): 
\begin{equation}
\mu (x,y)=1/2xy-x\theta (x-y)-y\theta (y-x)  \label{unn2}
\end{equation}

Regardless of the interpretation, the strategy in defining a self-adjoint
operator is the same: because the continuity is a property only of the
function itself but not of its derivatives, there will be surface terms that
will compete with lower dimensional integrals. They should be accurately
taken into account. That is exactly what we do below (and what we saw above
for the $n=2$ case).

Then we proceed exactly as we did in $n=2$ case: we will construct a
self-adjoint extension of the Hamiltonian, which in the `bulk', i.e.
everywhere except $x_{i}=x_{j}$ for all possible $i$ and $j$, reduces to a
Laplacian: 
\begin{equation}
-\Delta =-\sum_{i=1}^{N}\partial _{i}^{2}  \label{unn3}
\end{equation}
As usual, this will amount to imposing some sewing conditions on derivatives
of $f(\vec{x})$ at $x_{i}=x_{j}$. Let us start with the integration of the
highest dimensionality in (\ref{scalprod}), i.e. when all the clusters
contain just one coordinate, $X_{i}=x_{i}$. Performing integration by parts
and taking into account surface terms, we have 
\begin{align}
\int d^{N}\vec{x}\ g^{\ast }(\vec{x})\partial _{i}^{2}f(\vec{x})& =\int d^{N} \vec{x}\ (\partial _{i}^{2}g)^{\ast }(\vec{x})f(\vec{x})+  \notag \\
& -\sum_{j\neq i}\int \prod_{k\neq i}dx_{k}\ \left[ g^{\ast }(\vec{x})\partial _{i}f(\vec{x})-(\partial _{i}g)^{\ast }(\vec{x})f(\vec{x})\right]
_{x_{i}=x_{j}-\epsilon }^{x_{i}=x_{j}+\epsilon }  \label{unn4}
\end{align}
Summing over $i$, we have the result for the full Laplacian. It is clear
that the action of the Hamiltonian on $(n-1)$-components of the
aforementioned vector-function, i.e. when only two of the coordinates are
equal, should exactly compensate the unwanted term 
\begin{equation}
\sum_{i}\sum_{j\neq i}\int \prod_{k\neq i}dx_{k}\ \left[ (\partial
_{i}g)^{\ast }(\vec{x})f(\vec{x})\right] _{x_{i}=x_{j}-\epsilon
}^{x_{i}=x_{j}+\epsilon }  \label{unn5}
\end{equation}
and simultaneously reproduce the term 
\begin{equation}
-\sum_{i}\sum_{j\neq i}\int \prod_{k\neq i}dx_{k}\ \left[ g^{\ast }(\vec{x})\partial _{i}f(\vec{x})\right] _{x_{i}=x_{j}-\epsilon
}^{x_{i}=x_{j}+\epsilon }  \label{extraterm2}
\end{equation}
What is the Hamiltonian acting on $f_{ij}:=f(\vec{x})|_{x_{i}=x_{j}}$ that
does this? The answer is obvious, if we recall that the function $f$ itself
is continuous. Then we can define 
\begin{equation}
\hat{h}_{ij}f_{ij}=-\alpha \left( \partial _{i}f(\vec{x})-\partial _{j}f(\vec{x})\right) _{x_{i}=x_{j}-\epsilon }^{x_{i}=x_{j}+\epsilon }+\tilde{h}_{ij}\ .  \label{hij}
\end{equation}
Here $\alpha $ is a combinatorial coefficient expressed in terms of $C_{P}$
from (\ref{scalprod}) (in fact, it is not hard to see that it is \textit{always} the same as for $n=2$: $\alpha =2$, e.g., it is explicitly seen for $n=3$ from (\ref{3part}): $\alpha =\frac{6}{3}=2$); $\tilde{h}_{ij}$ is some
operator in $(n-1)$-dimensional space, such that its non-Hermiticity is on
the next, $(n-2)$-dimensional, level. In a moment we will see that it is
nothing but a Laplacian acting on $f_{ij}$. It is also easy to understand
why there are two terms in (\ref{hij}) instead of the one, as one would na\"{\i}vly expect from (\ref{extraterm2}): the term at $x_{i}=x_{j}$ will
arise twice in the sum - the first time as
\begin{equation}
\left[ g^{\ast }(\vec{x})\partial _{i}f(\vec{x})\right] _{x_{i}=x_{j}-\epsilon }^{x_{i}=x_{j}+\epsilon }  \label{unn6}
\end{equation}
and the second time as 
\begin{equation}
\left[ g^{\ast }(\vec{x})\partial _{j}f(\vec{x})\right] _{x_{j}=x_{i}-\epsilon }^{x_{j}=x_{i}+\epsilon }  \label{unn7}
\end{equation}

The residual freedom in the definition of $\hat{h}_{ij}$ coming from yet
undefined $\tilde{h}_{ij}$ cannot be fixed only by requiring Hermiticity - $\hat{h}_{ij}$ (and its analogs in lower dimensions) remain undetermined. It
is removed by requiring that it should agree with the matching conditions
that follow from the explicit action of the quantum Hamiltonian on the state
(\ref{wavefunct}). As we had already mentioned in the previous section, the
matching conditions, following from the action of the quantum Hamiltonian (\ref{Quant.Ham.}) on the $n$-particle state (\ref{wavefunct}), are,
essentially, the same as in the two-particle sector case (\ref{matc.cond1}).
Below we show that providing the matching conditions (\ref{matching2}) 
\begin{equation}
-\left( \partial _{j}f(\vec{x})-\partial _{i}f(\vec{x})\right)
_{x_{j}=x_{i}-\epsilon }^{x_{j}=x_{i}+\epsilon }=\partial _{j}\partial _{i}f(\vec{x})|_{x_{i}=x_{j}}\ ,\ \ \forall i>j  \label{matching}
\end{equation}
then $\hat{h}_{ij}f_{ij}$ is, in fact, equal to $\Delta f(\vec{x})|_{x_{i}=x_{j}}$ if $\tilde{h}_{ij}$ equals to ($n-1$)-dimensional Laplacian.

Let us find how a Laplacian in a lower dimensional space is related to the
one in $n$-dimensional evaluated when some coordinates coincide. For the
future use we consider more general case when there is one cluster but not
necessary of the size two, $\{x_{k_{1}},...,x_{k_{j}}\}$. As above, let $t$
be a collective coordinate for this cluster and $i$ runs over the rest of the
coordinates. Then the lower dimensional Laplacian is 
\begin{equation}
\tilde{\Delta}f_{k_{1}...k_{j}}(t,x_{i})=(\partial _{t}^{2}+\sum_{i}\partial
_{i}^{2})f(t,x_{i})  \label{unn8}
\end{equation}
Now for $\partial _{t}^{2}f(t,x_{i})$ we have 
\begin{equation}
\partial _{t}^{2}f(t,x_{i})=\left( \sum_{i:x_{i}\in \{x_{k}\}}\partial _{i}f(\vec{x})\right) _{\{x_{k}\}=t}+2\sum_{i>j:x_{i},x_{j}\in \{x_{k}\}}\partial
_{i}\partial _{j}f(\vec{x})|_{\{x_{k}\}=t}  \label{unn9}
\end{equation}
and finally 
\begin{equation}
\Delta f(\vec{x})|_{\{x_{k}\}=t}=\tilde{\Delta}f_{k_{1}...k_{j}}(t,x_{i})-2\sum_{i>j:x_{i},x_{j}\in \{x_{k}\}}\partial _{i}\partial _{j}f(\vec{x})|_{\{x_{k}\}=t}  \label{Delta}
\end{equation}

Now we see that if we define $\hat{h}_{ij}$ as 
\begin{equation}
\hat{h}_{ij}f_{ij}=-2\left( \partial _{i}f(\vec{x})-\partial _{j}f(\vec{x})\right) _{x_{i}=x_{j}-\epsilon }^{x_{i}=x_{j}+\epsilon }+\tilde{\Delta}_{i=j}f_{ij}  \label{unn10}
\end{equation}
and express $\tilde{\Delta}_{i=j}f_{ij}$ using (\ref{Delta}), we will have 
\begin{equation}
\hat{h}_{ij}f_{ij}=\Delta f(\vec{x})|_{i=j}  \label{unn11}
\end{equation}
which, of course, should be the case as Hamiltonian should take continuous
function to continuous function.

We can repeat each step starting at any level $(n-k)$ and using collective
coordinates, ${t_{i}}$. Because the non-Hermitian part on this level is in
corresponding $\tilde{h}$, which has exactly the form of $(n-1)$-dimensional
Laplacian, we will arrive at the level $(n-k-1)$ with exactly the same form
of the lower dimensional Hamiltonian. This completes the prove that the
Hamiltonian, given as a Laplacian in the `bulk' plus the matching condition (\ref{matching}), is, in fact, a self-adjoint operator in the space $L_{\mu }(R)$.

\section{S-matrix factorization}

Although the S-matrix factorization is the underlying property of quantum
integrable systems, it is quite difficult to analytically prove it using the standard perturbative calculations. Until now such calculations were fully
performed in all orders only for the non-linear Schr\"{o}dinger model \cite{Thacker:1974kv, Thacker:1976vp}, where the S-matrix factorization was
indeed proven - first for the three-particle scattering process, and then
generalized to the $n$-particle scattering process. Recently, in \cite{Melikyan:2008cy} we have considered the S-matrix factorization following
similar perturbative calculations, based on earlier two-particle S-matrix
calculations of \cite{Klose:2006dd}. We were able to show the three-particle
S-matrix factorization in the first non-trivial order in the perturbation
series. As we explained in \cite{Melikyan:2008cy}, there are essential
differences between the S-matrix calculations for the NLS and LL models.
Besides the technical difficulties associated with rapidly increasing, in
each perturbation order, number of vertices in the LL model, which leads to
a complex diagrammatic analysis, there are conceptual difficulties related
to the identification within the field theoretic approach of the Bethe
particles.

Here, however, we will establish the S-matrix factorization of the $su(1,1)$
LL model in a more direct fashion, following the results of the previous
sections, where we have explicitly constructed the self-adjoint extensions
and their continuity properties. Thus, having the exact expressions for the $n$-particle wave-functions and using the matching conditions, it is not
difficult to establish the S-matrix factorization. In fact, one can proceed
in the same fashion as for the simpler NLS model (see for example \cite{Korepin:1997bk}). Indeed, the main result from the previous section, that
allows explicit calculation of the $n$-particle S-matrix, is that the
matching conditions for the $n$-particle case (\ref{matching}) are exactly
the ones for the two-particle case (\ref{matc.cond1}). In other words, it is
enough to solve the equations for the two-particle case, in order to obtain
the solution for the general $n$-particle case. This is, essentially, the
S-matrix factorization.

To construct explicit expressions, we follow the derivation of \cite{Korepin:1997bk} for the NLS model. It is easy to see that for the LL model,
the matching condition (\ref{matching2}) leads to the following $n$-particle wave-function: 
\begin{equation}
f_{n}(x_{i}|p_{j})=(const)\underset{\{i\}}{\sum }(-1)^{\{i\}}e^{\sum \limits_{i=1}^{n}x_{i}p_{\{i\}}}\underset{i\neq j}{\prod }\left(
p_{\{i\}}-p_{\{j\}}-\frac{1}{2}ip_{\{i\}}p_{\{j\}}\right)
\label{wavefunction}
\end{equation}
where $\{i\}$ denotes all possible permutations of $(1,...,n).$ Thus, the $n$-particle wave-function is factorized in terms of the two-particle
wave-functions. As a consequence, the $n$-particle scattering S-matrix has
the form:
\begin{equation}
S_{n}(p_{1},...,p_{n)}=\underset{i\neq j}{\prod }S_{2}(p_{i},p_{j})
\label{Sn}
\end{equation}
where the two-particle scattering S-matrix for the $su(1,1)$ LL model has
the form: 
\begin{equation}
S_{2}=\frac{2(p_{1}-p_{2})-ip_{1}p_{2}}{2(p_{1}-p_{2})+ip_{1}p_{2}}
\label{S2}
\end{equation}
Let us note, that for the NLS model the expression for the wave-function has
similar to (\ref{wavefunction}) form, where the third term in the brackets
is a constant. Here, we have momenta product instead, which is the result of
derivatives present in the interaction vertex (see for details \cite{Klose:2006dd}). With this in mind, we can intuitively think of the LL model as the NLS model with momentum-dependent interaction.

\section{Conclusion}

We have considered the quantum integrable properties of the Landau-Lifshitz
model, and proposed a method to construct the quantum Hamiltonian. Most
importantly, we achieve this directly in the continuous case by regularizing
the ill-defined Hamiltonian, and constructing the necessary self-adjoint
extensions. This method allowed us to consistently derive the spectrum,
which we show to coincide with the one following from the quantum inverse
scattering method. We gave an explanation and derived in the most general $n$-particle case the puzzling construction of Sklyanin \cite{Sklyanin:1988s1}
(for the particular two-particle sector) of the quantum-mechanical
Hamiltonian action on the vector-like state. The continuity properties of
the functions involved in the construction of such states have also been
carefully investigated. These properties are defined by\ the corresponding
matching conditions, and, as we have shown, lead to the S-matrix
factorization property.

The particular difficulties of the LL model quantization are the ill-defined
operator product of the local conserved charges, as well as the highly
singular potential in the quantum-mechanical picture, which make it impossible
the use of the trace identities in the quantum case. Thus, it is clear, that the method considered in this paper should be applicable to any continuous integrable model which
has a singular nature. Since we have only considered the isotropic LL model,
the next natural problem is to consider the anisotropic LL model, which is
of great importance in the theory of integrable models. This, however, seems
to be a more complex task, as the algebra of observables should be, for
consistency, modified by hand in the inverse scattering method, forming the
Sklyanin algebra. Let us remind, that this algebra is naturally obtained
from the lattice models, but its appearance in the continuous models is less
clear. It would be interesting to give a direct derivation of the Sklyanin
algebra without appealing to the lattice version, since in more complex
continuous integrable models the construction of the corresponding lattice
models, as we have discussed in the introduction, is generally a quite
complex task, that has not been well-understood even for simple models.

Another important problem, also discussed in introduction, is to investigate
the $su(2)$ LL model. Let us remind that in the Sklyanin's original paper 
\cite{Sklyanin:1988s1} as well as in our work, only the quantization
of the $su(1,1)$ LL model is considered. This is done to have physically meaningful states
in the chosen representation for the states, consistent with the
ferromagnetic choice of the vacuum. Constructing representations in the $su(2)$ case with positively defined metric seems to be a more complex task
which has not so far been considered in connection with the quantization of the
LL model. It is known that such representations are possible to construct,
but the vacuum will not be ferromagnetic anymore. Thus, the algebraic Bethe
ansatz is not applicable in this case, and one has to consider more
sophisticated methods of finding algebraic solutions, much like for the
sinh-Gordon model \cite{Sklyanin:1989cf, Sklyanin:1989cg}.

Although we have considered only regularization and diagonalization of the
first non-trivial conserved charge (the Hamiltonian), integrability implies
the conservation of the infinite tower of charges, that should be possible
to regularize in the manner similar to the method proposed in this paper. We
do not currently know whether it is possible to do in a unified manner, or
each charge should be considered separately. Clearly, the questions posed
above for the LL mode, as the simplest representative of associated
difficulties, will appear in other more interesting continuous integrable
model. As an example, we mention the recently discovered
Alday-Arutyunov-Frolov fermionic model, which appears in the $su(1,1)$
subsector of the $AdS_{5}\times S^{5}$ strings \cite{Alday:2005jm}. There we
expect the similar difficulties to appear in the quantization process, as
the singular nature of the fermionic interaction terms will clearly require
construction of the self-adjoint extensions and careful consideration of the
conserved charges. These and other related problems are currently under
investigation.

\section*{Acknowledgment}

The work of A.M. was partially supported by the FAPESP grant No. 05/05147-3. The work of A.P. was partially supported by the FAPESP grant No. 06/56056-0. We would like to thank Brazilian Ministry of Science and Technology (MCT) and the Instituto Brasileiro de Energia e Materiais (IBEM) for the partial support of our research.

%\clearpage
\bibliographystyle{utphys}
\bibliography{ll_extensions}

\end{document}